\documentstyle [prl,aps,twocolumn,epsf, psfig]{revtex}
\newskip\humongous \humongous=0pt plus 1000pt minus 1000pt
\def\caja{\mathsurround=0pt}
\def\eqalign#1{\,\vcenter{\openup1\jot \caja
    \ialign{\strut \hfil$\displaystyle{##}$&$
    \displaystyle{{}##}$\hfil\crcr#1\crcr}}\,}
\newif\ifdtup

\relax 

\begin{document}

\newcommand{\newc}{\newcommand}

\newc{\be}{\begin{equation}}
\newc{\ee}{\end{equation}}
\newc{\ba}{\begin{eqnarray}}
\newc{\ea}{\end{eqnarray}}
\newc{\bea}{\begin{eqnarray}}
\newc{\eea}{\end{eqnarray}}
\newc{\D}{\partial}
\newc{\ie}{{\it i.e.} }
\newc{\eg}{{\it e.g.} }
\newc{\etc}{{\it etc.} }
\newc{\etal}{{\it et al.}}

\newc{\ra}{\rightarrow}
\newc{\lra}{\leftrightarrow}
\newc{\no}{Nielsen-Olesen }
\newc{\tp}{'t Hooft-Polyakov }
\newc{\lsim}{\buildrel{<}\over{\sim}}
\newc{\gsim}{\buildrel{>}\over{\sim}}

\title{Superconducting String Texture}
\bigskip
\author{L. Perivolaropoulos$^a$ and T.N. Tomaras$^b$} 
\address{$^a$ Institute of Nuclear Physics, N.C.R.P.S. Demokritos, GR-153
10, Athens, Greece \\ $^b$ Department of Physics and Institute of
Plasma Physics, University of 
Crete, and FO.R.T.H., \\ P.O. Box 2208, GR-710 03 Heraklion, Greece} 

\maketitle
\begin{abstract}
We present a detailed analytical and numerical study 
of a novel type of static, superconducting, classically stable 
string texture in a 
renormalizable topologically trivial massive $U(1)$ gauge model
with one charged and one neutral scalar. 
An upper bound on the mass of the charged scalar as well as
on the current that the string can carry are established.  
A preliminary unsuccesful 
search for stable solutions corresponding to
large superconducting loops is also reported. 
\end{abstract}

\noindent

The term texture is attributed generically to topological configurations 
trivial at spatial infinity. The winding of the fields takes
place over a finite region which roughly defines the location of 
the configuration.
Such defects have attracted considerable attention, both
in particle physics and cosmology. Well known examples are
the {\it skyrmion} which offers a useful alternative description
of the nucleon, and the {\it global texture} used recently to
implement an appealing mechanism for structure formation
in the Universe. In cosmological applications one makes use of
the instability of three dimensional texture in renormalizable
purely scalar theories. All such configurations are unstable 
towards shrinking, they collapse to a point and eventually decay
to scalar radiation. This is a natural decay mechanism, which
on the one hand prevents the domination of the energy density
by texture-like defects, and on the other it leads to highly
energetic events, which can provide the primordial fluctuations
necessary for structure formation. 

In particle physics one would be more interested
in observing such solitons in accelerator experiments, and
the above instability is an unwanted feature. One approach
to stabilize such configurations was the introduction into the
action of higher derivative terms. However,
being non-renormalizale, 
such terms are undesirable in the tree level action, and 
furthermore it has not been possible so far to produce 
in a controlable unambiguous way a quartic
term of the right sign to lead to stable solitons. 
An alternative way has been advocated recently and has
succesfully stabilized texture in realistic extensions of the
Standard Model with more than the minimal one Higgs doublet
content. The texture here is stabilized by the gauge interactions.

An extended Higgs sector in the effective low energy 
theory of electroweak interactions is favoured by 
supersymmetry, superstring theory and is necessary if
one wishes to arrange for an
efficient and potentially 
realistic electroweak baryogenesis.
Examples of simple realistic models with a multiple 
Higgs field content are the two Higgs-doublet standard model
(2HSM), and the minimal supersymmetric standard model (MSSM).
It is well known that no finite energy 
topological strings or particle-like
solitons exist in these models, and furthermore 
if no additional spontaneously broken discrete 
global symmetries are introduced, they do not carry domain
walls either.

However, it was pointed out recently \cite{bt1}, 
that an extended Higgs 
sector supports generically the existence of a new
class of quasi-topological metastable solutions.
Like topological solitons 
these objects are characterised 
by a winding, which counts the number of times the relative
phase of the Higgs multiplets winds around its manifold as we
scan the space transverse to the defect. 
Unlike topological solitons on the other hand, 
their existence is not decided by the symmetry structure alone
of the theory. In particular, they do not exist for all values 
of parameters and are at best classically stable. They are
local minima of the energy functional and decay to
the vacuum via quantum tunneling \cite{bt1}.
Alternatively, they can be thought of as embedded textures, 
which, in contrast 
to the previously discussed electroweak or Z-strings \cite{va}, 
are trivial at spatial infinity.
We occasionally 
refer to these defects generically as ${\it ribbons}$, 
reminiscent  
of the way they look in the simplest 1+1 dimensional 
paradigm presented in \cite{bt1}, \cite{bt}.

The above ideas have been explicitly demonstrated 
with the construction of membranes 
\cite{bt1} \cite{bt3} \cite{dtn} \cite{rt}
and infinite straight
strings \cite{bt2} \cite{brt}
in the context of simple toy models or in the 2HSM for realistic
values of its parameters, including the MSSM as a special case.
Finally, even though no stable particle-like solitons have been
suggested so far in any of these realistic models, 
a new tower of sphaleron solutions was obtained, characterised by
a finite number of modes of instability \cite{btt}. 
When they exist, these new solutions 
have lower energy than the standard model 
sphalerons or deformed sphalerons
and furthermore, they are less unstable having smaller in magnitude
eigenfrequencies of instability.

The string texture in particular discussed 
so far, may also be viewed as
semilocal three dimensional, static, classically stable 
generalizations of the
Belavin-Polyakov solitons \cite{bp} of the 
$O(3)$ non-linear $\sigma-$model.
A massive $U(1)$ \cite{bt2} 
or the $SU(2)\times U(1)$ \cite{brt} gauge fields of the
2HSM stabilize these solitons against the shrinking instability
induced by the scalar potential
terms \cite{derrick}.
The charged fields vanish at the center of the string, but are 
non-zero on a tube of radius and thickness both 
of electroweak scale, surrounding its axis.
This configuration is a novel kind of bosonic 
superconducting string. Contrary to the one 
presented in \cite{witten}, it is not absolutely stable, 
it has different topological characteristics and 
is more "economical" employing the same Higgs field 
both for its formation as well as for its
superconductivity.

Two issues arise naturally. First, to what extent 
is it possible
to generalise the above stable string solutions and allow 
for a current to flow along them, 
while retaining their stability? 
More importantly, could there exist stable particle-like 
configurations, current-carrying loops of such superconducting
strings?
Being of the electroweak scale, such a loop would correspond 
to a particle with mass of a few 
TeV and would be the first example of a stable 
soliton in a realistic model of particle physics with a chance
to be produced in the next generation of accelerator 
experiments.

Clearly, it should not be surprising that one may
in principle allow for a current to flow along
the string. 
After all, a perturbatively
small current may reduce slightly the stability of the string,
but it should not make a local minimum of the energy
dissappear altogether.

The existence of stable loops
on the other hand, depends crucially on the maximum 
current such a string can sustain.
Imagine a piece of length $2\pi R$ 
of superconducting string with 
thickness $\bar\rho$ and winding $Q$ in the 
transverse directions.  
Introduce a current along it by a twist $\nu \equiv N/R$
of $N$ full turns of the phase of the charged scalar 
over the string length $2\pi R$, and glue the ends of the string
together to form a loop. Parametrize by $\varphi$ the 
angle around the loop and by $\rho, \theta$ the radial 
coordinate and the polar angle 
in the plane transverse to the string.
The configuration may then be 
represented by $\Phi = f(\rho) e^{iQ\theta} e^{iN \varphi}$. 
The profile $f(\rho)$ may conveniently be approximated 
by a constant $\bar f$ on a tube 
surrounding the string axis. 
With $\tilde g$ the gauge coupling, 
the corresponding current density components are 
$J_{\varphi} \sim {\tilde g} {\bar f}^2 N / R$ flowing along the loop, 
and $J_{\theta} \sim {\tilde g} {\bar f}^2 Q/ \bar\rho$ perpendicular
to $J_{\varphi}$ roughly on a tube of radius $\bar\rho$ 
surrounding the string
axis. The total current $I$ circulating in the loop
is given by the surface integral of $J_{\varphi}$ over the
string cross section and equals 
$I \sim J_{\varphi} \pi {\bar\rho}^2 \sim {\tilde g} {\bar f}^2 N \pi \bar\rho^2 / R$.
Similarly, the current per unit string length in the $\theta$ direction
is $i \sim J_{\theta} \bar\rho \sim {\tilde g} {\bar f}^2 Q$. 
$I$ gives rise to a magnetic field whose flux through 
the superconducting loop is constant and given by  
$\Phi_0 \sim I R \sim {\tilde g}N {\bar f}^2 \pi {\bar\rho}^2$. Its 
energy is, up to inessential logarithmic corrections 
\cite{pc}, equal to
$E_m \sim \Phi_0^2/2R \sim ({\tilde g} {\bar f}^2 N \pi \bar\rho^2)^2/2R$. 
The string 
tension may be approximated by the magnetic energy
of the field produced by $J_{\theta}$. It is given by
$E_T \sim B_{\varphi}^2 (Volume)/2 \sim i^2 2\pi R \pi \bar\rho^2/2
\sim ( {\tilde g} {\bar f}^2 Q \pi \bar\rho)^2 R$.
The minimum of the total energy $E=E_m+E_T$ is at
$R / \bar\rho \simeq N /\sqrt{2} Q$, 
or equivalently at the value of the $\Phi$ twist
\be
\nu \simeq \sqrt{2} \; {Q \over \bar\rho}
\label{loop}
\ee
The pressure due to the squeezed magnetic field through 
the loop opposes the tendency of the loop to contract to
zero radius, and the system reaches an equilibrium with
radius given in (\ref{loop}). 

The argument of the preceding paragraph is based on
several simplifying assumptions. The string was treated
as a perfect superconducting wire, with definite 
thickness and perfect Meissner effect, while the
loop was assumed to have $R \gg \bar\rho$. 
However, the above discussion shows that it is unlikely to
form a stable loop the way we describe it here, unless
the straight string can support a current 
strong enough to satisfy (\ref{loop}).

The precise evaluation of the maximum current that 
a straight string texture can support and 
the existence of stable loops are 
dynamical questions, which require 
detailed numerical study.
In this paper we take a first step and examine these issues
in the context of a simple massive $U(1)$ 
gauge model \cite{bt2}, which captures most
of the relevant features of the 2HSM.
In section 1 we describe the model we shall
be interested in. A perturbative semiclassical analysis
is presented, which leads to the necessary and 
sufficient conditions for the existence of stable texture,
carrying the current induced by a fixed twist per
unit length in the charged scalar. Section 2 contains 
the detailed numerical study of the model. We confirm the
analytical results, we make precise the meaning of the 
conditions for stability obtained in section 1, and show
that (\ref{loop}) cannot be satisfied in the context
of this model. This is in line with the results of a first preliminary 
attempt to find stable loops, also reported
in section 2. 
A summary and some remarks concerning superconducting string texture
in the realistic 2HSM are offered in the discussion section.
Finally, a semiclassical proof that a massless $U(1)$ gauge field
does not lead to stable texture is presented in the Appendix.

\vspace{1cm}



\section{The model -  Semiclassical analysis}

\vspace{0.3cm}

A simple field theoretical laboratory \cite{bt2} to study the main features 
of string texture contains a complex scalar field 
$\Phi = \Phi_1 + i \Phi_2$ coupled to a massive U(1) gauge field $Z_\mu$ 
as well as to a neutral scalar $\Phi_3$. 
Their dynamics is described by the Lagrangian density
\begin{equation}
\eqalign{
{\cal L}=&{1 \over 2} (D_\mu \Phi)^\dagger D^\mu \Phi +
{1\over 2} \partial_\mu \Phi_3 \partial^\mu \Phi_3 - V(\Phi, \Phi_3) \cr
& -{1\over 4} Z_{\mu \nu} Z^{\mu \nu} + {1\over 2} m^2 Z_\mu Z^\mu
\label{model1}
}
\end{equation}
where $Z_{\mu\nu}=\partial_\mu Z_\nu-
\partial_\nu Z_\mu$ and
$D_\mu = \partial_\mu + i g Z_\mu$.
The gauge boson should be massive for stable
strings to exist (see Appendix I). We choose to call it Z 
because its role in the context of (\ref{model1}) is analogous to
that of $Z^0$ in realistic electroweak theories \cite{brt}.

The potential is given by
\begin{equation}
V(\Phi, \Phi_3)={\lambda \over 4}(\sum_{a=1}^3 \Phi_a^2 - v^2)^2
+{\kappa^2 \over 8}(\Phi_3 -v)^4+{1\over2}\mu^2|\Phi|^2
\label{potential1}
\end{equation}
The classical vacuum of the model is 
\be
\Phi=0 \;,\;\; \Phi_3=v
\label{vacuum}
\ee
and the masses of $Z$, $\Phi$ and $\Phi_3$ are $m$, $\mu$ and 
$m_H\equiv\sqrt{2\lambda} v$, respectively.
We have not considered the most general potential consistent with the
O(2) invariane of the model, nor have we tried to generate the gauge
boson mass more naturally via Higgs mechanism with an extra complex 
scalar. For convenience
we keep the number of fields and the couplings to a minimum.
As mentioned in the introduction,   
string texture of the type studied below has already 
been predicted to exist
also in a large class of realistic models \cite{brt}.
Of course, a $U(1)$ gauge field with an explicit mass term
does not spoil renormalizability, provided it couples to a
conserved current.

The field equations of the model are
\be
\partial^\mu Z_{\mu\nu} + m^2 Z_\nu = J_\nu
\label{Zequation}
\ee
\be
D^\mu D_\mu \Phi = -{{\partial V}\over{\partial {\Phi^\star}}}\;, \;\;\;
\partial^\mu \partial_\mu \Phi_3 = - {{\partial V}\over{\partial {\Phi_3}}}
\label{Phiequations}
\ee
The gauge current
\be
J_\mu \equiv {g\over {2i}} \Bigl(\Phi^\star D_\mu\Phi - 
(D_\mu\Phi)^\star \Phi\Bigr)
\label{current1}
\ee
is automatically conserved by the $\Phi-$equations of motion.  
Combined with (\ref{Zequation}) it implies the transversality
\be
\partial^\mu Z_\mu = 0
\label{constraint1}
\ee
of the gauge field.

Finally, the energy density of (\ref{model1}) is
\be
\eqalign{
{\cal E}=& {1 \over 2} [(D_0 \Phi)^\dagger D_0 \Phi +
(D_i \Phi)^\dagger D_i \Phi +
\partial_i \Phi_3 \partial_i \Phi_3] + V(\Phi, \Phi_3) \cr
& + {1\over 2} Z_{0 i} Z_{0 i} + {1\over 4} Z_{i j} Z_{i j} + 
{m^2 \over 2} (Z_0 Z_0 + Z_i Z_i)
\label{energy 1}
}
\ee
where $i,j = 1,2,3$.

Having a unique classical vacuum (\ref{vacuum}) 
and a trivial target
space the model does not support the existence of any kind of 
absolutely stable topological solitons. However, notice that 
in the naive limit
\be
\lambda \to \infty \;\;{\rm and} \;\; g, \kappa, \mu \to 0
\label{naive}
\ee
the magnitude $F \equiv \sqrt{\Phi_a \Phi_a}$ 
of the triplet $\Phi_a$ freezes at its vacuum value
$v$, and (\ref{model1}) reduces to a decoupled massive gauge field
plus the ungauged O(3) non-linear 
$\sigma-$model
\be
{\cal L}_0 = {{m_H^2}\over{2\lambda}} {1\over 2} \partial_\mu n^a \partial^\mu n^a
\label{L0}
\ee
for the unit-vector field 
\be
n^a \equiv {{\Phi_a} \over F}
\label{unitfield}
\ee

It is well known that ${\cal L}_0$ has topologically stable 
static string solutions \cite{bp}.
To simplify their description, one may replace the unit field $n^a$ 
by a complex scalar $\Omega$ through the stereographic projection
\be
n_1+in_2 = {{2\Omega}\over{1+|\Omega|^2}}\; , \; 
n_3 = {{1-|\Omega|^2} \over {1+|\Omega|^2}}
\label{bps1}
\ee
from the unit sphere $S^2$ onto the complex plane. 
The strings of (\ref{L0}) stretching along the $x_3$ axis,
are given by holomorphic functions
$\Omega(z)$, where $z=x_1+ix_2$. They are classified by the
number of times $Q$ the transverse 
two-space wraps around the target space. Convenient expressions
for this integer winding number
$Q$ are 
\be
\eqalign{
Q =& {1\over \pi} \int dx_1 dx_2 {{{\bar \partial}{\bar \Omega} \partial \Omega - 
{\bar \partial}\Omega \partial{\bar \Omega}} \over {(1+|\Omega|^2)^2 }} \cr
=& {1\over{8\pi}} \int dx_1 dx_2 
\epsilon_{\alpha\beta} \epsilon_{abc} n^a \partial_\alpha n^b
\partial_\beta n^c
\label{winding}
}
\ee
with $\partial \equiv \partial/\partial z$, and lowercase Greek
indices taking the values 1, 2 in the transverse directions. 
The simplest solution \footnote{A constant $w_0$ cannot be added to $\Omega_0$. 
Its energy per unit length would diverge quadratically for
non-vanishing $\kappa$, in which we shall be interested shortly.}
\be
\Omega_0 = {{\bar\rho e^{i\alpha}}\over{z-z_0}}
\label{bps}
\ee
with arbitrary constant $\bar\rho$, $\alpha$ and $z_0$, 
the only one that will interest us explicitly in this paper, 
describes an infinite string of "thickness" $\bar\rho$
stretched parallel to the third
axis through $z_0$; it has $Q = 1$ and 
energy per unit length $E_0=2\pi m_H^2 /\lambda$. 

It is natural to expect, that even if we should 
relax "slightly" the above limits on
the parameters, solutions close to (\ref{bps}) 
will continue to exist and be stable. 
Any statement about existence and classical stability of solutions should
of course depend only upon the classically 
relevant parameters of the model. Of the six parameters in
${\cal L}$, we choose $m$ to set the scale and define $m=1$. By appropriate
rescallings a second one may be pulled outside of the
action to play the role of 
the semiclassical parameter $\hbar$, and we are left with four
classically relevant ones. 
We rescale $F \to F/\sqrt{2\lambda}$ 
and $Z_\mu \to Z_\mu/\sqrt{2\lambda}$, to bring (\ref{model1}) to
the form
\be
\eqalign{
{\cal L} = {1\over{2\lambda}} \Bigl[& \; {1 \over 2} (\partial_\mu F)^2
+{1\over 2} F^2 |(\partial_\mu +i{\tilde g }Z_\mu)(n_1+in_2)|^2 \cr
&+{1\over 2} F^2 (\partial_\mu n_3)^2 - {1\over 8} (F^2-m_H^2)^2 \cr
&- {{{\tilde \kappa}^2}\over 8} (Fn_3-m_H)^4  
-{{\mu^2}\over2} F^2 (n_1^2+n_2^2) \cr
&- {1\over 4} Z_{\mu\nu}^2 + 
{1\over 2} Z_\mu Z^\mu \Bigr]
\label{rescaled} 
}
\ee
with the four classically relevant dimensionless parameters 
explicitly shown to be
\be
\mu \;, \;\; m_H\equiv \sqrt{2\lambda} v\;,\;\; 
\tilde g  \equiv {g\over\sqrt{2\lambda}}\;, \;\;
\tilde \kappa \equiv {\kappa\over{\sqrt{2\lambda}}}
\ee

Following \cite{bt2}, to find static minima of the 
energy we proceed in two steps.
First, we keep the unit vector field ${\bf n}$ fixed and
time independent, and minimize the energy with respect to 
the Higgs magnitude $F$ and the gauge field $Z_\mu$. 
Assuming they stay close to their vacuum values one finds:
\be
F\simeq m_H \Bigl[ 1-{1\over{m_H^2}} ((\partial_i {\bf n})^2+\mu^2 (n_1^2+n_2^2)) \Bigr]
\label{Fofn}
\ee
\be
Z_0=0 \;\; {\rm and} \;\; Z_k \simeq 2{\tilde g } m_H^2 \int d^3y G_{kl}(x-y) j_l(y)
\label{Zofn}
\ee
where
\be
j_l(x)={1\over 2}(n_2 \partial_l n_1 - n_1 \partial_l n_2)
\ee
and $G_{kl}(x-y)$ is the three-dimensional massive Green function
\be
G_{kl}(x)=\int {{d^3p}\over{(2\pi)^3}} e^{-i{\bf p}\cdot{\bf x}} 
{{\delta_{kl}+p_kp_l}\over{{\bf p}^2+1}} 
\ee

Using (\ref{Fofn}) and (\ref{Zofn}) one next 
eliminates $F$ and $Z_\mu$ from the energy
and is left with the effective energy functional for the angular field ${\bf n}$:
\be
\eqalign{
&E={m_H^2 \over {2\lambda}} \Bigl[\int d^3x {1\over 2} (\partial_i {\bf n})^2 \cr
 & + \int d^3x \Bigl({{\mu^2}\over2} (n_1^2+n_2^2)
 - {1\over{2m_H^2}} (\partial_i {\bf n} \partial_i {\bf n})^2 \\
+ {{\tilde\kappa}^2\over 8} m_H^2 (n_3-1)^4 \Bigr)  \cr
& - 2{\tilde g }^2 m_H^2 \int d^3x \int d^3y j_i(x) G_{ik}(x-y) j_k(y) \Bigr]
\label{Eofn}
}  
\ee
The first integral is the non-linear sigma model leading contribution. 
The terms in the second integral are the corrections due to the 
potential, while the last term is due to the gauge interaction. 
Our semiclassical perturbation scheme is consistent
provided 
\be
|F-m_H| \ll m_H \;\; {\rm and} \;\;
{\tilde g } Z_i {\bf n} \ll \partial_i {\bf n}
\label{conditions0}
\ee
are satisfied everywhere. 

The configurations ${\bf n}({\bf x})$ of interest in this 
article are current-carrying infinite strings, which may also 
be thought of as almost straight 
pieces of a large closed loop.
They will be taken of the form 
\be
n_1+in_2 = e^{i\nu x_3} \Bigl({\tilde n}_1(x_1, x_2)+i{\tilde n}_2(x_1, x_2)\Bigr)
\label{ansatz0}
\ee
with constant $\nu$. This is not the most general ansatz for
such a string, since $\nu$ could in general
depend also upon the transverse coordinates;
nevertheless it is expected to capture its 
main features \cite{vs94}.

For string configurations of the form (\ref{ansatz0}), with
thickness $\bar\rho$ in the transverse $(x_1,x_2)$ plane,
conditions (\ref{conditions0}) translate into:
\be
{1\over{m_H\bar\rho}}\;,\;\; {\nu \over {m_H}}\;,\;\; {\mu\over{m_H}}\;,\;\; 
{\tilde \kappa} m_H \bar\rho \;,\;\; 
{\tilde g } m_H min(\bar\rho, 1) 
\ll 1
\label{conditions1}
\ee
The thickness $\bar\rho$ will be determined dynamically
in the sequel, and one should a posteriori verify that the 
above constraints can indeed be satisfied.
Notice that contrary to what the naive limit (\ref{naive})
seems to suggest,
$\lambda$ does not have to be very large for
the validity of our conclusions. It may be arbitrarily
small, consistent with our semiclassical treatment, 
and still satisfy the
conditions of existence and stability of solutions, 
which are expressed in
terms of ${\tilde g }$, ${\tilde \kappa}$, $\mu$ and $m_H$.

To leading order in our approximation the model at hand
has the Belavin-Polyakov topological string solutions,
the simplest of which is configuration 
(\ref{ansatz0}) with $\nu=0$ and 
${\tilde n}_1 +i {\tilde n}_2$ given by (\ref{bps1}), (\ref{bps})
with arbitrary thickness $\bar\rho$. 
According to \cite{bt2}, turning on the interactions and, by the 
same reasoning, introducing a fixed twist per unit length as 
in (\ref{ansatz0}), affect to leading
order only the thickness $\bar\rho$ of the string.
To determine the position
of possible equilibrium values of $\bar\rho$ one should insert
into (\ref{Eofn}) 
the "twisted $Q=1$ Belavin-Polyakov" configuration
(\ref{ansatz0}) with ${\tilde n}_1 +i {\tilde n}_2$ 
given by (\ref{bps1}) and (\ref{bps}), 
and minimize 
the resulting expression of the energy per unit length
with respect to $\bar\rho$ 
\footnote{Translational and rotational invariance
imply that the energy of (\ref{ansatz0}) 
does not depend on $\alpha$ or $z_0$.}.

Consistency of our approximation requires the additional condition   
\be
\bar\rho \delta \equiv \bar\rho \sqrt{\nu^2+\mu^2} \ll 1
\label{conditions2}
\ee
and the energy per unit length takes the form:
\be
\eqalign{
{\cal E}(\bar\rho) \simeq &{{2\pi m_H^2} \over \lambda}\Bigl[ 1+ 
\delta^2 \bar\rho^2 ln({R\over{\bar\rho}}) 
+{1\over 6}{\tilde \kappa}^2 m_H^2 \bar\rho^2  \cr 
& -{8\over{3m_H^2 \bar\rho^2}} 
-{\tilde g }^2 m_H^2 \bar\rho^2 
\int_0^\infty dx {{x^3 K_0^2(x)}\over{x^2+\bar\rho^2}}
\Bigr]
\label{Eofr}
}
\ee 
$R$ is an infrared cut-off assumed to be much larger than $\bar\rho$.

A few comments are in order: 
First, the logarithmic divergence in (\ref{Eofr}) appears 
only in the case $Q=1$ studied here. It is due to the slow fall-off
at infinity of the $Q=1$ Belavin-Polyakov configuration, and 
dissappears for all higher $Q$. But even for $Q=1$ its presence
in (\ref{Eofr}) is an artifact
of our approximation. With non-vanishing $\nu$ and/or $\mu$
all fields approach their vacuum asymptotic values much faster
and all dangerous integrals become finite.
As will be verified numerically, no infrared divergence
is actually present in the energy 
and for all practical purposes $A\equiv ln(R/\bar\rho)$ 
should be interpreted as a constant
of order one. 
Second, notice that to the order of our approximation 
the current and the $\Phi$
mass enter in ${\cal E}(\bar\rho)$ only in the combination $\nu^2+\mu^2$,
and consequently they have the same effect on the zeroth order solution.
Finally, conditions (\ref{conditions1}) and (\ref{conditions2}), necessary
for the consistency of our semiclassical approach, may be
combined into
\be
{1\over{m_H \bar\rho}}\;,\;\; 
{\tilde\kappa} m_H \bar\rho\;,\;\; 
{\tilde g} m_H min(\bar\rho, 1)\; ,\;\; \bar\rho \delta \ll 1
\label{final}
\ee

According to (\ref{Eofr}), the twist, the $\Phi$-mass, and the potential, all tend
to reduce the string thickness. The gauge interaction tends to blow it up.  
Is it possible to obtain a stable equilibrium?
Following \cite{bt2}, where the case $\delta=0$ was analysed, 
we define $\Delta^2\equiv 6A\delta^2/m_H^2$, 
and conclude that for values of the parameters
\be
a\equiv {{{\tilde\kappa}^2+\Delta^2}\over{\tilde g }^2} \;\; {\rm and} \;\;
b\equiv {2\over{{\tilde g }^2 m_H^4}}
\label{ab}
\ee 
below the solid line of Figure 1 

\begin{figure}
\psfig{figure=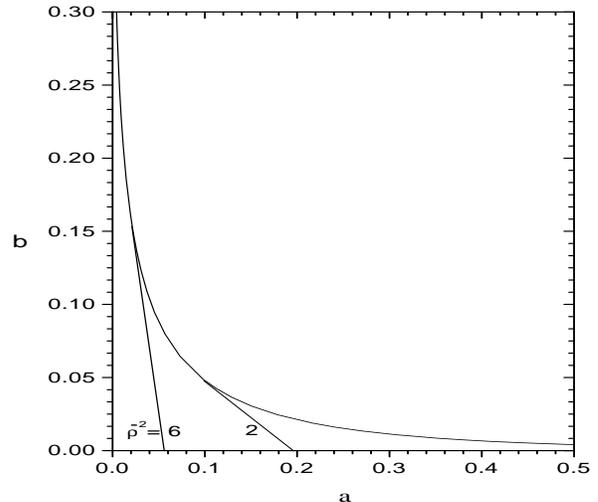,height=3.2in,width=9cm,angle=0} \caption{The 
semiclassical boundary 
of the stability region  for stable strings in the $(a,b)$ plane. The
square $\bar\rho^2$ of the thickess of some solutions is also shown. 
} \label{fig:qq0} 
\end{figure}

\noindent
and for small enough $1/m_H$ and $\delta$ to satisfy conditions 
(\ref{final}), a stable solution 
exists. For a given $\nu$ it is a small
deformation of the twisted $Q=1$ Belavin-Polyakov
configuration with size $\bar\rho$ as 
shown on the corresponding
tangent to the curve. Its energy per unit length is
guaranteed by (\ref{final}) to 
differ only slightly from $E_0=2\pi m_H^2/\lambda$.  
The precise meaning of inequalities (\ref{final}),
as well as 
the computation of the lower bound on $m_H$ and of the 
upper bound on $\delta$ for a stable solution to exist 
corresponding to a given
point $(a,b)$
in the stability region are dynamical questions dealt with 
numerically in the following sections.

\vspace{1cm}



\section{Numerical results}
 
\vspace{0.3cm}

\subsection{String texture}

In this subsection we shall perform a detailed numerical study of 
the string texture solutions 
of (\ref{model1}) in order to verify and extend the analytical 
semiclassical results
reviewed briefly above. We find it convenient to start
with $\delta=0$ and leave the more general case for a later section.

\vspace{0.3cm}

{\it The ansatz}

We use the most general static ($\partial / \partial t = 0$), 
$x_3-$independent ($\partial / \partial {x_3} =0$), 
axially symmetric ansatz for an infinite straight string with 
winding $Q$ stretched 
along the $x_3-$axis
\be
\eqalign{
\Phi = f(\rho) e^{i Q \theta} \; &, \;\; \Phi_3 = G(\rho) \cr
{\bf Z} = {\bf e}_\theta \; & K(\rho) 
\label{stringansatz}
}
\ee
with $\rho$ and $\theta$ the usual polar coordinates in the 
transverse plane.  
For static configurations the $Z_0$ dependent part of the energy 
density is the sum  
$(\partial_i Z_0)^2 + m^2 Z_0^2 + e^2 |\Phi|^2 Z_0^2 $
of three positive terms   
minimized for $Z_0(\rho)=0$. Similarly, $Z_3(\rho)=0$ 
and $Z_{\rho}(\rho)=0$. 
Note that constraint (\ref{constraint1}) 
and current conservation are automatically
satisfied by the ansatz. 

The energy density and the current of the ansatz
in terms of the rescaled quantities, for which we keep the
same symbols, are
\be
\eqalign{
{\cal E} &= {1 \over {2 \lambda}}
\Bigl[ 
{1\over 2}(K^\prime + {K \over \rho})^2 + 
 {1\over 2} f^{\prime 2} + 
{1\over 2} ({Q \over \rho} - {\tilde g } K)^2 f^2 + {1\over 2} K^2 \cr 
& + {1\over 2} G^{\prime 2} + 
{1 \over 4} (f^2 + G^2 - m_H^2)^2 + {{\tilde \kappa}^2 \over 8} (G-m_H)^4 
\Bigr]
\label{energy3}
}
\ee
and
\be
J_\theta = -{\tilde g } \Bigl({Q\over \rho} - 
{\tilde g } K(\rho) \Bigr) f^2
\label{J}
\ee
respectively, while the magnetic field 
${\bf B}\equiv \nabla \times {\bf Z}$ points in the 3-direction
and is equal to
\be
B_3 = K^\prime + {{K(\rho)}\over \rho}
\label{B}
\ee 

Extremizing the energy functional one is led to the following 
field equations for the unknown functions $f$, $G$ and $K$:
\be
\eqalign{
&-(K' + {K \over \rho})' - 
{\tilde g } ({{Q}\over \rho} -{\tilde g } K)f^2 + K =0 \cr 
&-{1\over \rho} (\rho f^\prime)' + ({Q \over \rho} - 
{\tilde g } K)^2 f + (f^2 + G^2 - m_H^2) f = 0 \cr
&-{1\over \rho} (\rho G^\prime)' + (f^2 + G^2 - m_H^2) G + 
{{\tilde \kappa}^2 \over 2} (G-m_H)^3 = 0
\label{fieldequations}
} 
\ee
It may be checked that they coincide with equations 
(\ref{Zequation}) and (\ref{Phiequations}) evaluated
for the ansatz.

\vspace{0.3cm}

{\it The boundary conditions} 

As usual, finiteness of the energy and the field 
equations are used to determine the boundary conditions. 
It is straightforward to check that in the present 
case of vanishing $\Phi$-mass
and twist, the solution at infinity 
behaves like
\be
\eqalign{
f(\rho) \sim C_1/\rho^{Q} \;, &\;\; 
G(\rho) \sim m_H-C_1^2/2m_H\rho^{2Q} \cr
K(\rho) \sim &{\tilde g }Q C_1^2/\rho^{2Q + 1}
\label{bc1}
}
\ee
while specifically for $Q=1$, the case of interest below, 
its behaviour at the origin is 
\be
f(\rho) \sim C_2 \rho \;,\;\; 
G(\rho) \sim C_3 + C_4 \rho^2 \;, \;\;
K(\rho) \sim {\tilde g} C_2^2 \rho
\label{bc2}
\ee
with constant $C_i$, $i$=1,2,3 and $C_4$ related to $C_3$ 
by $8C_4+(m_H-C_3)[2C_3(C_3+m_H)+{\tilde\kappa}^2(C_3-m_H)^2]=0$. 
Consequently, the energy density 
of a $Q=1$ string 
behaves as $1/\rho^4$ at large distances.


\vspace{0.3cm}

{\it Numerics - Solution search}

To search for string texture solutions of (\ref{model1})
we used a relaxation method \cite{press} 
with locally variable mesh size
and the convenient set of boundary conditions  
\be
f(0)=0 \;, \;\; K(0)=0 \;, \;\; (\rho f^\prime)(0)=0 
\ee
\be
G(\infty)=m_H \;, \;\; (\rho G^\prime)(\infty)=0 \;,\;\; B_3(\infty)=0
\label{bc}
\ee
following from (\ref{bc1}) and (\ref{bc2}). One starts with an initial
trial configuration, which is iteratively improved 
until it becomes a solution
of the field equations within satisfactory accuracy. 
As an extra check of the 
accuracy of the solutions obtained, 
we used three virial conditions, whose general form
we shall describe in the next section.
Typically they were satisfied within one part in $10^3-10^4$. 
Finally, to make sure that the solutions correspond to local
minima of the energy and are stable, 
we perturbed slightly each one of
them, using a large number of smooth random perturbations and
verified that the perturbed configurations were always 
of higher energy.

As explained in the previous section, stable solutions are 
not expected to exist in an arbitrary model (\ref{model1}),
but only in those with parameters
within the stability region. Using the 
semiclassical results to guide the search, 
one starts with a choice of $(a,b)$ in the stability 
region. The tangent
to the thick curve that passes through 
$(a,b)$ corresponds to
a size value $\bar\rho(a,b)$. According to the 
semiclassical analysis
a stable solution should exist, which is a small
deformation of (\ref{bps}) with size $\bar\rho(a,b)$, provided all
conditions (\ref{final}) are satisfied. Figure 1 shows that
$\bar\rho$ lies typically
between one and five, while $a$ and $b$ are smaller than one. 
Thus, to satisfy the third constraint in (\ref{final})
\be
{\tilde g } m_H = \sqrt{{2\over b}} {1\over {m_H}} \ll 1
\label{emH}
\ee
one should take
\be
m_H \gg \sqrt{{2 \over b}}
\label{mH}
\ee
All remaining conditions are then automatically satisfied.
A general remark which follows from the semiclassical
analysis is that models with parameters in the upper left
corner of Figure 1 
favour the existence of thick strings, 
with the constraints satisfied for relatively 
low Higgs masses. On the other hand, to find 
thin strings, one has to search in models with large 
Higgs mass, and parameters in the lower right corner of
Figure 1.

To summarize, the theory with a given set of values of 
$(a,b)$ in the stability region, and
$m_H$ satisfying (\ref{mH}), should have a stable solution
close to (\ref{bps}) with size 
$\bar\rho(a,b)$. The values of the couplings ${\tilde g }$
and ${\tilde \kappa}$ follow from $a$, $b$ and $m_H$.
Accordingly, a good guess for the initial configuration necessary
for our numerical procedure is configuration
(\ref{bps}) for the scalars and vanishing gauge fields.

\vspace{0.5cm}

{\it Results}

We start with the verification 
that stable solutions exist. We restrict ourselves 
throughout to
the most interesting case $Q=1$.

Applying the recipe of the previous paragraph, 
choose $a$=0.001, $b$=0.2 and $m_H=4$. They correspond to
${\tilde g }\simeq 0.2$, ${\tilde \kappa}\simeq 0.006$. 
The profile of the stable texture obtained
with an initial configuration with $\bar\rho=6.7$ is shown
in Figure 2. We have been able to go deeper inside the
upper left corner of Figure 1 and find stable string texture 
for $m_H$ as low as two.

\begin{figure}
\psfig{figure=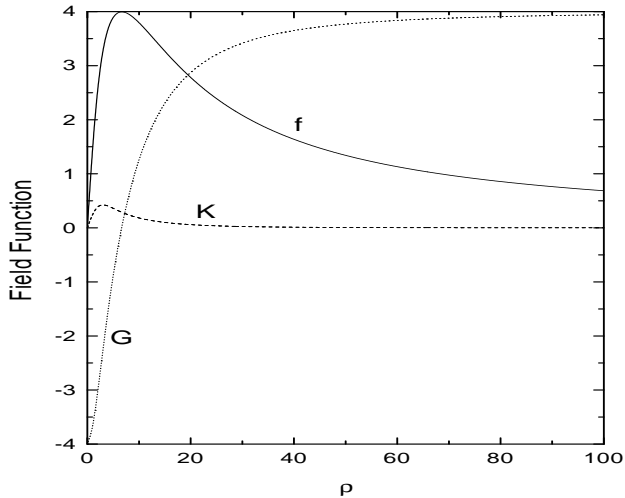,height=3.2in,width=9cm,angle=0} \caption{The 
profile of the string in the model with ${\tilde g }=0.2$, 
${\tilde \kappa}=0.006$ and $m_H=4$. Its energy is $E=13.6\times 
4\pi$. } \label{fig2} 
\end{figure}

Similarly, Figure 3 presents the profile of the solution 
for $a$=0.25, $b$=0.01 and $m_H=20$. It corresponds to
the model with ${\tilde g }=0.04$ and ${\tilde \kappa}=0.02$.

\begin{figure}
\psfig{figure=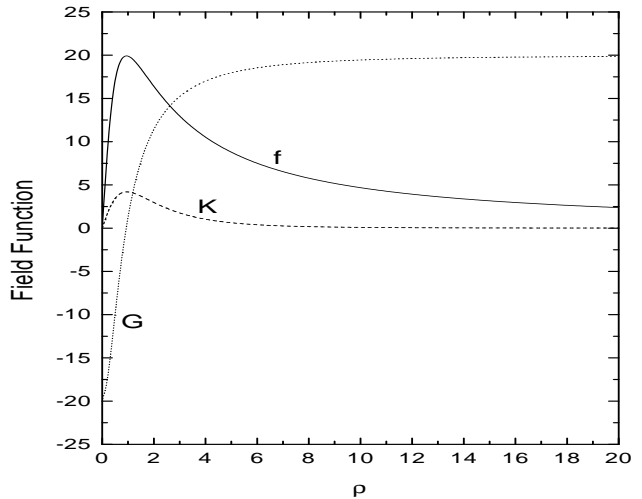,height=3.2in,width=9cm,angle=0} \caption{For 
parameters in the lower right corner of Figure 1 stable solutions 
require larger $m_H$ and are thinner. } \label{fig3} 
\end{figure}

For both solutions presented above the value of 
${\tilde g } m_H \simeq 0.8$. Thus, the constraint 
(\ref{emH}) should in practice be interpreted roughly as
${\tilde g } m_H < 1$. 
Notice that like in the wall case \cite{bt3} 
all string solutions discussed here have energies per unit length
smaller and within 20\% from the 
value $4\pi m_H^2$ \footnote{Energies in our numerics are defined
up to the overall 
factor $1/2\lambda$ in (\ref{energy3}).} corresponding to the 
limiting Belavin - Polyakov solution.

In Figure 4 we plot the Higgs magnitude $F(\rho)\equiv 
\sqrt{f^2+g^2}$, the magnetic field $B_3$ and the current 
$J_\theta$ for the second solution. 
\begin{figure}
\psfig{figure=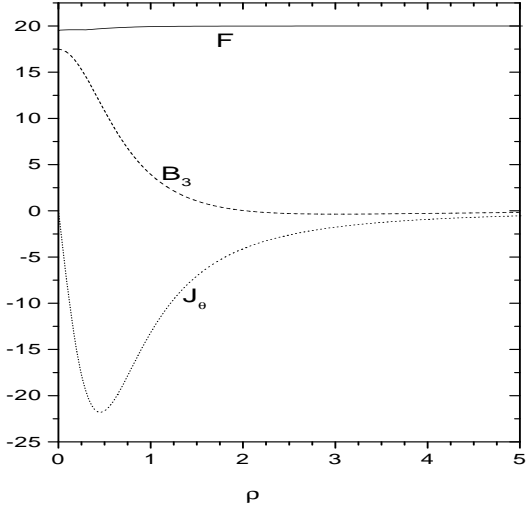,height=3.2in,width=9cm,angle=0} \caption{The 
profiles of the Higgs magnitude (solid line), the magnetic field 
(dashed) and the current (dotted) of the string texture of Figure 
3. } \label{fig4} 
\end{figure}

Note that the Higgs magnitude differs, in accordance with the 
theoretical analysis, only slightly from its vacuum value $m_H$. 
Furthermore, it is everywhere non-zero, so that the unit vector field 
$n_a\equiv \Phi_a / F$ is well defined and the corresponding 
winding number (\ref{winding}) unambiguous. Finally, it should
be pointed out that the 
magnetic field takes both positive and negative values. One 
may verify that the total magnetic flux is zero, as expected from 
the asymptotic behaviour of the gauge field in (\ref{bc1}). 


For fixed values of $a$=0.02 and $b$=0.05 we find solutions
for a variety of $m_H = 10, 20, 30, 50$. 
Their sizes (defined approximatelly for the purposes
of this plot by the zero of 
$G(\rho)$) are plotted in Figure 5 
against $m_H$ and shown to be roughly constant in
accordance with the semiclassical analysis.

For very large $m_H$ though one expects deviations from this
result. According to Appendix I the thickness of the 
solutions should eventually increase with $m_H$ and for
very large Higgs mass be pushed to infinite size.
No stable string exists for zero gauge boson mass.

Our next task is to perform a numerical study of the extent of the 
stability region in the $(a,b)$ plane and compare it against the 
semiclassical result. We were unable to find stable texture for 
parameters $(a,b)$ lying above the dashed curve in Figure 6. Notice 
the remarkable agreement with the leading order semiclassical 
curve also depicted for convenience by the solid line. 

\begin{figure}
\psfig{figure=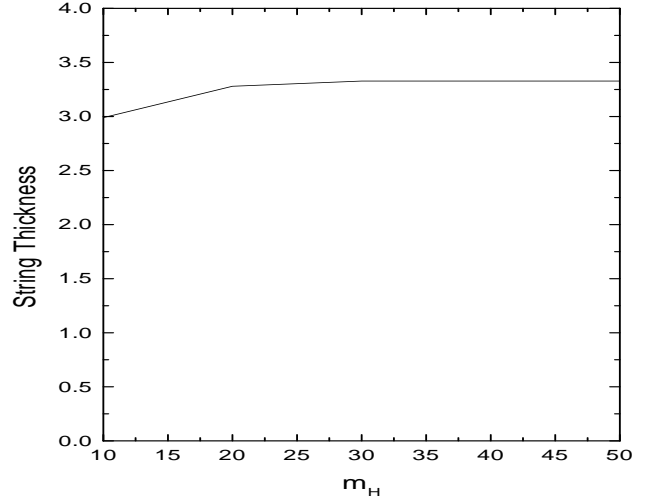,height=3.2in,width=9cm,angle=0} \caption{ 
For fixed $a$ and $b$ the thickness of the string is rather 
insensitive to the value of $m_H$. } \label{fig5} 
\end{figure}

\begin{figure}
\psfig{figure=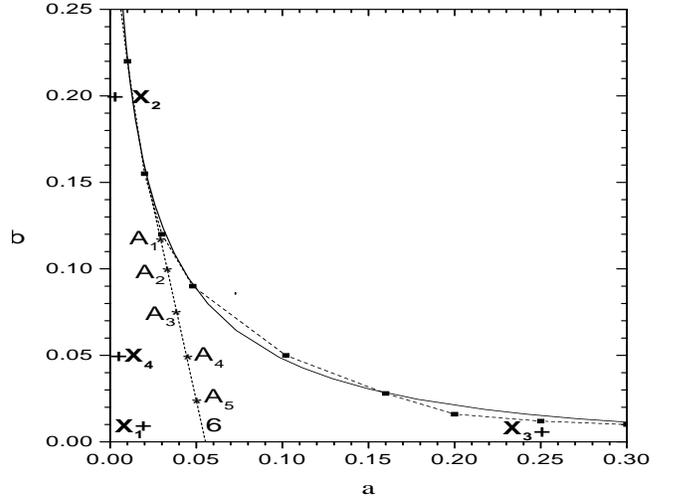,height=3.2in,width=9cm,angle=0} \caption{ 
 The stability region as determined numerically
(dashed line) plotted together with the semiclassical result 
(solid curve).} \label{fig6} 
\end{figure}

Finally, it is interesting to test the semiclassical prediction 
that an infinite set of theories, characterized by parameters on a 
line of fixed $\bar\rho$, all lead to string solutions of the same 
thickness. The sizes of the solutions obtained for the theories 
corresponding to the points $A_1$ to $A_5$ on the line of Figure 6 
corresponding to $\bar\rho \simeq \sqrt{6}$ are plotted in Figure 
7. The Higgs mass was chosen in such a way that the quantity 
${\tilde g } m_H $ is constant and equal to 0.5. 

\begin{figure}
\psfig{figure=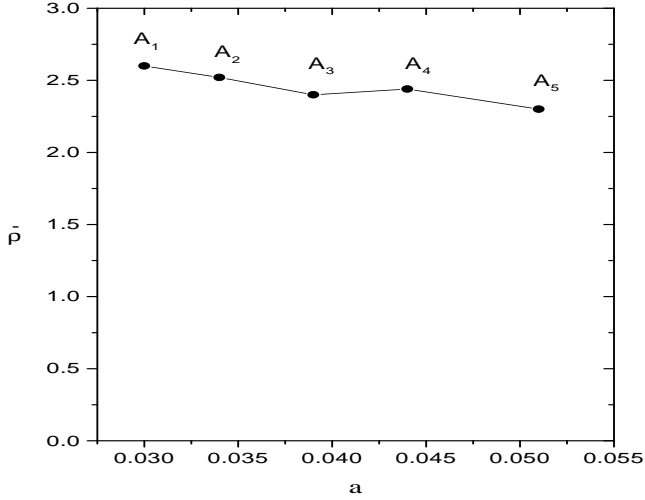,height=3.2in,width=9cm,angle=0} \caption{ 
 The sizes of the string solutions for parameter
values corresponding to the points $A_1$ to $A_5$ of Figure 6, 
plotted against $a$.} \label{fig7} 
\end{figure}

\subsection{Twisted strings - Wire quality}

Next, we shall extend the previous results and 
search numerically for current-carrying string
texture.
We take the string, preferably thought of as 
a long straight piece of a large
loop, stretching along 
the $x_3$-axis and generalise the axially symmetric
ansatz used in the previous section, 
to include a twist in the complex scalar along $x_3$  
\begin{equation}
\eqalign{
&\Phi = f(\rho) e^{i Q \theta}  e^{i u(x_3)}\, , \;\; \Phi_3 = G(\rho) \cr
&{\bf Z} = {\bf e}_\theta \;K(\rho) + {\bf e}_3 \;W(\rho)
\label{string-ansatz}
}
\end{equation}

The gauge current flowing along the string is given by 
$ J_3 = -g(du/dx_3 - g W(\rho)) f^2 $. Its conservation 
translates into 
$
{{d^2 u(x_3)}/{dx_3^2}} = 0
$
that is, to a linear dependence of the phase $u(x_3)$ upon $x_3$.
We shall take the scalar phase to make $N$ full turns over
the length $2\pi R$ of the string, and set the constant term to zero. 
This fixes 
\be
u(x_3) = {N\over R} x_3 \equiv \nu x_3
\ee

In terms of the rescalled dimensionless fields 
and coordinates 
defined in the previous section 
and conveniently denoted by the same symbols,
the energy density of the ansatz is:
\be
\eqalign{
{\cal E} &= {1\over {2 \lambda}}
\Bigl[ 
{1\over 2}(K^\prime + {K \over \rho})^2 + 
{1\over 2} W^{\prime 2} + {1\over 2} f^{\prime 2} + {1\over 
2} ({Q \over \rho} - {\tilde g } K)^2 f^2 
\cr 
& + {1\over 2} (\nu - {\tilde g } W)^2 f^2 + 
{1\over 2} G^{\prime 2} + {1 \over 4} 
(f^2 + G^2 - m_H^2)^2 \cr
&+ {{\tilde \kappa}^2 \over 8} (G-m_H)^4 
+ {1\over 2} (K^2 + W^2)
\Bigr]
\label{energy4}
}
\ee
Correspondingly, the $x_3$-component of the gauge current becomes
\be
J_3 =-{\tilde g }\Bigl(\nu - {\tilde g } W(\rho) \Bigr) f^2 
\label{J3}
\ee

Extremizing the energy (\ref{energy4}) 
one obtains the field equations
\be
-(K' + {K \over \rho})' - 
{\tilde g } ({{Q}\over \rho} -{\tilde g } K)f^2 + K = 0
\label{fieldequations1}
\ee
\be
-{1\over \rho} (\rho W')' - 
{\tilde g } ( \nu - {\tilde g } W)f^2 + W = 0
\label{fieldequations2}
\ee 
\be
\eqalign{
-{1\over \rho} (\rho f')' + ({Q \over \rho} - {\tilde 
e} K)^2 f +&( \nu - {\tilde g } W)^2 f \cr
+ &(f^2 + G^2 - m_H^2) f = 0
\label{fieldequations3}
}
\ee
\be
\eqalign{
-{1\over \rho} (\rho G')' + (f^2 +& G^2 - m_H^2) G  \cr
+& {{\tilde \kappa}^2 \over 2} (G-m_H)^3 = 0
\label{fieldequations4}
}
\ee
which we shall solve numerically for fixed non-zero $\nu$, 
following the same approach as in the previous section.

\vspace{0.3cm}

{\it The boundary conditions.} 

Finiteness of the energy forces the 
configuration to tend to the vacuum at spatial infinity
\footnote{For a large circular loop the center of
the loop is also a point at infinity.}. 
A convenient 
set of conditions there is given by (\ref{bc2}) together with
\be
W(\infty)=0
\ee
At the center on the other hand we keep (\ref{bc1}) and add
\be
(\rho W')(0)=0
\ee
for $W(\rho)$.

Configuration (\ref{string-ansatz}), viewed as 
a circular loop and with the above boundary conditions 
which effectively compactify space into $S^3$, 
defines a map from $S^3$ onto $S^2$, the target
space of the unit-vector field (\ref{unitfield}), which 
as explained before is well defined for all solutions of
interest in this paper. 
As such it is characterised by the Hopf topological index 
\be
H = {1\over {8\pi}} \int_{S^2} \epsilon_{abc} \Phi_a d\Phi_b d\Phi_c 
= Q \cdot N
\label{hopf}
\ee
Notice that for $Q=1$ one may interpret $\nu/2\pi$ as the Hopf
charge per unit string length.

\vspace{0.3cm}

{\it An upper bound on the twist magnitude.}
 
It is instructive to view the twisted string as a small deformation
of the untwisted one.
For $\nu=0$ the $W$-equation gives $W=0$ and the problem reduces to the
untwisted case discussed in the previous section. Consider 
such a stable string corresponding to parameters inside the stability
region and to a value of $m_H$ not exceeding say 20, to stay near the 
phenomenologically interesting regime.
Start increasing $\nu$, while keeping $\tilde\kappa$,
$\tilde g$ and $m_H$ fixed. During this process $b$ in (\ref{ab})
stays fixed, while $a$ increases. Eventually, at some critical
value $\nu_C$, one will cross
the solid curve of Figure 1 and the string solution will 
disappear altogether \footnote{Note the difference from the 
phenomenon of {\it current quenching} observed previously 
in the context of superconducting strings \cite{witten} 
with topological stability.
Contrary to the latter case, not only the current but the 
string itself disappears to radiation 
once we exceed the critical value of
the twist.}. 
$\nu_C$ depends on the values of the other parameters.
To maximize the current one should arrange for the maximum relevant
value $a_{max}$ of $a$ within the stability region of Figure 1. 
This corresponds to the lowest value of $b$, which 
as a consequence of (\ref{mH}) cannot for $m_H < 20$ exceed 
the value $b_{min} \simeq 0.01$. Figure 1, then leads to an 
$a_{max} \simeq 0.3$, which according to (\ref{ab})
translates into ${\Delta}_{max}^2 / {\tilde g}^2 \simeq 0.3$. 
Combined with the constraint (\ref{emH}) on the value of
${\tilde g} m_H$ we obtain
\be
\delta_{max} \simeq 0.2
\label{delta}
\ee
Thus, the maximum current one may hope to drive through such
a string corresponds to a twist $\nu_{max}=0.2$. 
Similarly, according to (\ref{delta}) the value 0.2 
is also an estimate of the upper bound 
on the charged Higgs mass, consistent 
with the existence of stable strings. 
String texture corresponding to $m_H \geq 20$ may 
of course support stronger currents and allow for 
larger $\mu$.
In any case, given that according to our analysis, 
the effects of non-zero $\mu$ and $\nu$
are identical 
to a high degree of accuracy, we set
$\mu=0$ throughout the numerical study that follows.

\vspace{0.3cm}

{\it Virial relations.} 

Three virial conditions were used to check
the accuracy of the solutions discussed in this paper. 
They express the 
stationarity of the energy functional under particular 
deformations of the solution. 
By the standard argument, imagine a solution of the field equations was found. 
It is an extremum of the energy. Any small change of the configuration
should have to linear order the same energy as the original one.
The derivative of the energy functional with respect to the parameters 
parametrizing the deformation should vanish when evaluated at the
solution.
The virial conditions we used are
\be
E_1 - 2 E_2 = 0
\ee
\be
E_1 - 2 E_3 = 0
\ee
and
\be
2 E_4 - E_5 = 0
\ee
relating
\be
E_1 = {\pi\over\lambda}\int_0^L d\rho \; Q \; {\tilde g } \; f^2 \; K
\ee
\be
\eqalign{
E_2 = &{\pi\over {2\lambda}} 
\int_0 ^L d\rho \;  \rho \Bigl[ K^2 (1+{\tilde g }^2
f^2)+{1\over 2} (f^2 + G^2 - m_H^2)^2 \cr
&+ {{\tilde \kappa}^2 \over 4} (G - m_H)^4 +
W^2 + (\nu - {\tilde g }\; W)^2 f^2 \Bigr]
}
\ee
\be
E_3 ={\pi\over {2\lambda}}\int_0^L d\rho \; \rho 
\Bigl[ (K' + {K\over \rho})^2 + (1+ {\tilde g }^2 f^2)\;  K^2 \Bigr]
\ee
\be
E_4 = {\pi\over {4\lambda}}\int_0^L d\rho \; \rho f^4
\ee
and
\be
\eqalign{
E_5 = {\pi\over {2\lambda}} \int_0^L & d\rho \; \rho \Bigl[ {f^\prime}^2 +
({{Q}\over \rho}-\tilde g  K)^2 f^2  \cr 
& + (\nu-\tilde g  W)^2 f^2 +
(G^2 - m_H^2) f^2 \Bigr]
}
\ee
They arise by demanding stationarity of the energy with respect to
solution size rescalling 
$\rho \to \alpha \rho$, 
$K$-rescalling $K(\rho) \to \beta K(\rho)$ 
and $f$-rescalling
$f(\rho) \to \gamma f(\rho)$,
respectively. Such field rescallings are consistent, as they
ought to,
with the boundary conditions on the fields $K$ and $f$.

All solutions obtained numerically satisfied the
above virial conditions to a very good approximation. 
Specifically, in all
cases the appropriately normalized virial
quantities $v_1=|(E_1-2E_2)/(E_1+2E_2)|$, $v_2=|(E_1-2E_3)/(E_1+2E_3)|$ and
$v_3=|(2E_4-E_5)/(2E_4+E_5)|$ were of ${\cal O}(10^{-4} - 10^{-3})$.

\vspace{0.3cm}

{\it Results.}

To find twisted solutions we start with an untwisted one 
as initial trial configuration, and iteratively improve it
until it becomes a solution of 
(\ref{fieldequations1})-(\ref{fieldequations4}) with
the given value of $\nu$. 
  
Figure 8 shows the profile of the solution arising 
by the above method from the untwisted string corresponding to
point $X_1$ with $a=0.02$ and $b=0.01$ in Figure 6. 
For the remaining parameters we chose $m_H=20$ and $\nu=0.05$.  
As for the initial ansatz we took the Belavin-Polyakov 
soliton with $\bar\rho=3.5$ and vanishing gauge fields. 

\begin{figure}
\psfig{figure=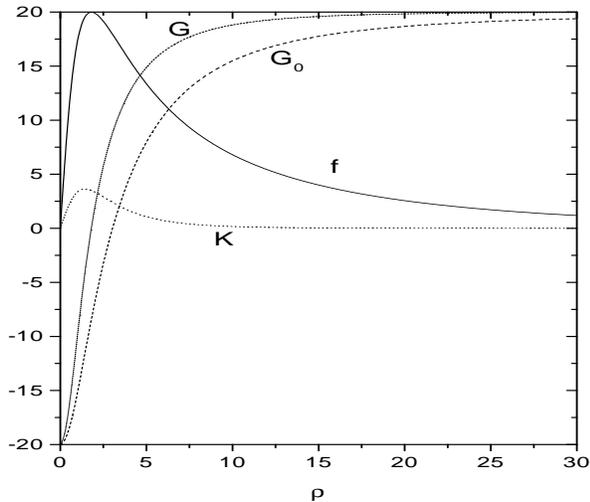,height=3.2in,width=9cm,angle=0} \caption{ 
The twisted string for ${\tilde g}=0.035$, $\tilde \kappa = 
0.005$, $m_H=20$ and $\nu=0.05$. On the same plot we also show the 
profile of the function $G_0(\rho)$ of the untwisted solution. The 
twist reduces string thickness.} \label{fig8} 
\end{figure}

To observe the destabilization of the string solution caused by a 
large current and to determine the value $\nu_{max}$ of the twist, 
we continued increasing $\nu$ for fixed values of the remaining 
parameters. For the solution corresponding to $X_1$ presented in 
Figure 8 we found $\nu_{C1} \simeq 0.1$. In a similar fashion, we 
computed the maximum currents supported by the untwisted string 
textures plotted in Figures 2 and 3, whose corresponding $(a,b)$ 
are shown in Figure 6 by the points $X_2$, $X_3$ and $X_4$. The 
maximum values of the twist found are $\nu_{C2} \simeq 0.01$, 
$\nu_{C3} \simeq 0.04 
 \; {\rm and} \; \nu_{C4} \simeq 0.04$, respectively.
The agreement of these results with the semiclassical absolute 
bound (\ref{delta}) obtained above is rather satisfactory. The 
corresponding total current, roughly equal to $I \sim \pi 
\bar\rho^2 {\tilde g} \nu m_H^2$, is less sensitive. It was
evaluated numerically and shown to take values
between 30 and 52 for the above solutions. The most 
promising region of parameters for the existence of 
stable closed loops is around 
$X_1$, but still $\nu$ cannot easily become large enough to 
satisfy (\ref{loop}).

\vspace{1cm}



\subsection{Large string loops}

An interesting question, that needs to be addressed in the context of 
our toy model, is the question of spring 
formation \cite{witten}, \cite{chht88}. 
The analysis so far does not allow much hope that stable string
loops can exist in (\ref{model1}). The semiclassical prediction 
(\ref{delta}) or even worse
the numerically determined maximum value of the twist are much smaller
than the value (\ref{loop}) required for spring formation. In fact
the last semiclassical constraint in (\ref{final}), even when interpreted
as a simple inequality as suggested by all numerical results obtained 
above, leaves little room if at all for stable loops. 
Furthermore, one should note that (\ref{loop}) is rather optimistic 
for our toy model, because it was obtained for massless gauge 
field which maximizes the
magnetic pressure due to the trapped magnetic flux. 

In any case, proper numerical search for string loops in this model would then
mean to look for rather small loops with inner radius of the order
of the gauge field inverse mass or less, in order to maximize the effect
of the gauge field against loop contraction.  
This requires essentially full three dimensional analysis 
and was left for a future publication.

However, within the numerical approximation used in this paper,
we did verify the above conclusions for large loops of
radius $R \gg \bar\rho$. We approximated the loop by a straight string
of length $2\pi R$ and looked for minima of the energy
\be
E(R) = \int_0^{2\pi R} dz \int_0^R d\rho \, \rho \, {\cal E}(f,G,K,W)
\label{totenergy}
\ee
to check whether the $R$-dependent term $(N/R - {\tilde g} W)^2 f^2$
in the integrand (\ref{energy4}), which for fixed $N$ acts against
loop contraction, might actually stabilize it at some $R=R_{spring}$.    

Clearly, keeping $N$ constant, spring formation could occur only 
for $R$ large enough so that the solution exists \ie $N/R = \nu < 
\nu_C \equiv N/R_C$ corresponding to the chosen values of the 
parameters. If this minimum of the energy could be achieved at 
some $R_{spring} > R_C$ then at the $R_C (m_H)$, $E(R)$ would have 
a negative derivative with respect to $R$ i.e. the total energy $E$ 
would tend to decrease towards its minimum as $R$ increased from 
$R_C$ towards $R_{spring}$. 
%
%
We have checked all points at $R_C$s for a wide range of 
parameters ($4  < m_H < 1000$, $ 0 < {\tilde g} < 0.5$, $0<{\tilde 
\kappa} < 0.5$) in regions where solutions exist. We focused on 
regions where $R_C$ could be minimized (large $m_H$) thus 
maximizing the twist induced pressure of the $R$-dependent term 
$(N/R - {\tilde g} W)^2 f^2$. It is this term that could 
potentially stabilize the closed loop. We found that $(dE / 
dR)|_{R=R_C} > 0$ at all points with practically no signs of 
change even at the smallest $R_C$'s. Therefore, in line with the 
previous discussion, we conclude that for the parameter sectors we 
investigated no spring solutions exist. 

\vspace{1cm}

\section{Discussion}

To summarize, 
we have found stable current-carrying vortex solutions
in gauged generalizations of the $O(3)$ non-linear $\sigma$-model,
with a single $U(1)$ gauge field and the usual scalar triplet.
The model considered is an extension of that studied in
\cite{bt2}, and may also be viewed as semilocal \cite{va}.
Indeed, it has
generically a trivial vacuum manifold, while its target space
should effectively be thought of as an $S^2$ with an
$S^1$ gauged by the $U(1)$ gauge field. 
In this model we have mapped the parameter
sectors where stable solutions exist, while no stabilized
spring solutions were found in the parameter sectors discussed. 
The parameter region
corresponding to stable untwisted string texture \cite{bt2} has also been
examined and we confirmed numerically the approximate semi-analytical
results of that analysis.
An alternative way to stabilize vortex loops is the introduction of angular
momentum whose conservation can stabilize loops against collapse more
effectively than twist pressure. Loops stabilized by angular momentum are
known as {\it vortons} \cite{vortons} in order to be distinguished from
springs.

It is instructive at this point to examine what the
above results, obtained in the context of the toy model (\ref{model1}),
suggest about the two Higgs-doublet standard
model. As mentioned before, the gauge field in 
(\ref{model1}) corresponds to the
$Z^0$ gauge boson, while the role of $\Phi$ is here played by
the charged Higgs $H^+$.   
Clearly, the numerical results of the present paper 
strengthen our confidence to the semiclassical conclusions 
reported in \cite{brt}, which 
should be valid with high accuracy. In addition, the constraints
are weeker and should be interpreted as simple inequalities.  
Thus, the 2HSM supports stable strings. They may be 
generalized and allow
for a current to flow along them. The current 
due to the twist of the electrically charged 
$H^+$ is a bona-fide electric current
and the string texture in this case is
a superconducting wire in the standard sense. It is 
characterized by the twist parameter $\nu$, the Hopf
charge per unit string length.
Being of electroweak scale these "wires" should have a 
thickness of a few $m_W^{-1}$ and mass density
of the order of $10^{-4} gr/cm$. 
Extrapolating naively to the 2HSM the bounds 
obtained above, one is led to a maximum current they can carry of
about $10^8-10^{10}$Amp${\grave e}$res, 
corresponding to $\nu = \nu_{max} \sim 0.2$.
Equivalently, these bounds would imply   
the absence of stable string texture for $H^+$ mass 
$\mu$ larger than $\mu_{max} \sim 0.2 m_Z \sim 18 GeV/c^2$. 
Since this value is lower than the experimental lower bound on the
$H^+$ mass, little space is left for stable string texture
in the 2HSM for realistic values of its parameters. 

But, this last conclusion may well be too naive. The presence in 
general of a separate coupling for electromagnetism 
and of a richer variety of charged and neutral fields,
will change the maximum current
allowed along the string, as well as condition
(\ref{loop}), derived for the case of the single 
gauge field of the toy model studied in the present paper. 
What actually happens in more complicated models like the 2HSM 
is a matter of detailed analysis and deserves further study.

\vspace{2cm}

{\bf ACKNOWLEDGMENTS}
This work is partly the result of a network supported by the European Science
Foundation (ESF).
The ESF acts as a catalyst for the development of
science by bringing together leading scientists and funding agencies to
debate, plan and implement pan-European initiatives.
This work was also supported by the EU grant CHRX-CT94-0621, as well
as by the Greek General Secretariat of Research and Technology grant
$\Pi$ENE$\Delta$95-1759.

\vspace{2cm}

{\bf APPENDIX I}

\vspace{0.3cm}

{\it A massless U(1) gauge field does not lead to stable texture}.  
A massless gauge field is either too efficient in halting 
the shrinking caused by the potential terms and 
blows-up the texture to infinite thickness, or it is not efficient
and the string contracts to vanishing cross section. 

Here we sketch a semiclassical proof valid for
thick strings. More generally, the statement has been 
verified numerically.
It was shown in the main text that the introduction of
either mass or twist to the charged scalar 
works against the stability
of the string texture. It suffices to prove the statement 
for massless charged scalar and vanishing twist.
Start from (\ref{model1}) with $m=\mu=0$. 
Define the Higgs mass 
$\sqrt{2\lambda} v = 1$ to set the mass scale and rescale
fields and distances according to:
\be
F \to v F \;,\;\; A_\mu \to v A_\mu \;, \;\; 
x^\mu \to x^\mu / \sqrt{2\lambda} v
\ee
after which the action is written as
\be
\eqalign{
&{\cal L}_{m=0}={v\over{\sqrt{2\lambda}}} \Bigl[ 
{1\over2} (\partial_\mu F)^2 + 
{1\over2} F^2 |(\partial_\mu +i{\tilde g} A_\mu)(n_1+in_2)|^2 \cr
+& {1\over2} F^2 (\partial_\mu n_3)^2 
- {1\over 8}(F^2-1)^2 - 
{{\tilde \kappa}^2 \over 8} (Fn_3 - 1)^4 - 
{1\over 4} F_{\mu\nu}^2 \Bigr]
\label{Lm=0}
}
\ee
with fields and parameters defined as in the main text. Here
we have only changed to $A_\mu$ the name of the gauge field.

In the limit
\be 
{\tilde g}, {\tilde\kappa} \to 0
\ee
the model has absolutely stable topological strings (\ref{bps}) with
$F=1$. What will happen to such a soliton of arbitrary 
size $\bar\rho$ if we
move slightly away from the limit? Switching-on the potential
term will tend to shrink it, while a non-vanishing gauge coupling
will tend to blow it up. 

Following the steps of \cite{bt2} it is straightforward 
to solve for the magnitude $F$ and the gauge field, and derive 
an effective action for the unit-vector field $n_a$. 
Under the constraints
\be
\bar\rho \gg 1 \;, \;\; {\tilde\kappa}\bar\rho \ll 1 \;,\;\; \tilde g \ll 1
\ee
the energy per unit length is written as 
${\cal E} = E_0 + \delta {\cal E}$ with 
\be
E_0={v\over \sqrt{2\lambda}} \int d^2x {1\over 2} (\partial_i n_a)^2
\ee
and 
\be
\eqalign{
\delta {\cal E}={v\over \sqrt{2\lambda}} 
\Bigl[&-{1\over2}\int d^2x (\partial_i n_a \partial_i n_a)^2 + 
{{\tilde\kappa}^2 \over 8} \int d^2x (n_3-1)^4 \cr 
&-{\tilde g}^2 \int d^2x \int d^2y J_i(x) G_{ij}(x,y) J_j(y) \Bigr]
\label{dE}
}
\ee
The current is $J_i\equiv {1\over2}(n_2\partial_i n_1-n_1\partial_i n_2)$
and the Green function of the massless gauge field is 
\be
G_{kl}(x)=\int {{d^2p}\over {4\pi^2}} e^{-i{\bf p}\cdot{\bf x}} 
{{\delta_{kl}+p_k p_l}\over p^2} 
\ee

Following the same steps as in the main text, we 
evaluate ${\cal E}$ for the solution (\ref{bps}),
minimum of the leading term $E_0$. 
The result is
\be
{\cal E}_{m=0}(\rho)={v\over \sqrt{2\lambda}} 4\pi 
\Bigl[ 1 - {8\over{3\bar\rho^2}} + 
{1\over {12}} (2{\tilde\kappa}^2-3{\tilde g}^2)\bar\rho^2 \Bigr]
\label{Em=0}
\ee
The constant is the leading Belavin-Polyakov value for
the $Q=1$ soliton. The remaining terms represent the
leading correction to its energy due to the potential and 
the gauge interaction. 

This function does not have a local minimum. Q.E.D.

We should like to point out that this result is quite general
in our approximation. 
Dimensional analysis alone
fixes the gauge contribution to the energy to be 
$ \sim Constant \times \bar\rho^2$. It is "Lenz" that fixes 
the coefficient of the quartic
term in (\ref{dE}) to be negative. Relaxing the
constraint on $F$ reduces the energy of the configuration. 
Thus, for any value of $Q$ the energy takes the form
$\delta E \sim 1 - C_1 / \bar\rho^2 + C_2 \bar\rho^2$ with $C_1 > 0$.
Independently of the value of $C_2$ this function has no
local minimum.

For $\Delta \equiv 3{\tilde g}^2 - 2{\tilde\kappa}^2 < 0$ the gauge 
repulsion is not strong enough to halt shrinking to zero size. 
For $\Delta > 0$ the energy has a local maximum at
$\bar\rho_0 \equiv (4\sqrt{2} / (3{\tilde g}^2-2{\tilde \kappa}^2))^{1/4}$.
Strings of thickness smaller than $\bar\rho_0$ shrink to zero, while
those of initial thickness larger than $\bar\rho_0$ blow-up to
infinity. 

\vspace{1cm}



\vfill

\eject
\end{document}